\documentclass{article}
\usepackage{graphicx,amssymb,amsmath}

\begin{document}

\title{Study of the $D^0 \rightarrow \pi^+ \pi^- \pi^0$ decay at BABAR}


\author{
  Mario Gaspero on behalf of the BABAR Collaboration  \\ \\
  \em {Dipartimento di Fisica, Sapienza Universit\`a di Roma, and }\\
 \em {Istituto Nazionale di Fisica Nucleare, Sezione di Roma 1 }\\
 \em {Piazzale Aldo Moro 2, I-00185, Rome, Italy } \\
 \texttt{mario.gaspero@roma1.infn.it}
}

\date{}

\maketitle



\begin{abstract}
The Dalitz-plot of the decay $D^0 \rightarrow \pi^+ \pi^- \pi^0$ measured by 
the BABAR collaboration shows the structure of a final state having quantum
numbers $I^G J^{PC} = 0^- 0^{--}$\@.
An isospin analysis of this Dalitz-plot finds that the fraction of the $I=0$ 
contribution is about 96\%\@.
This high $I=0$ contribution is unexpected because the weak interaction 
violates the isospin.
\end{abstract}

\section{Introduction}

\mbox{~~~~This communication} reports the results of the analysis of the 
Cabibbo suppressed decay\footnote{
Charge conjugate decay modes are implicitely included.}
\begin{equation}
D^0 \rightarrow \pi^+ \pi^- \pi^0
\label{D03pi}
\end{equation}
made by the BABAR collaboration.

The BABAR detector \cite{detector} measured the $e^+ e^-$ 
annihilations at the PEP-II collider.
Most of the data were taken at the $\Upsilon(4S)$ resonance to study the $B$
mesons.
The detector measured also the decays of the charm mesons and baryons
generated by the decay of the $B$ mesons or by the {\em continuum\/}, i.e.\
by the $e^+ e^-$ annihilations into $q \bar{q}$.

The charged tracks were measured by a {\em silicon vertex tracker\/} and
by a {\em drift chamber\/}, and identified by a {\em ring-imaging Cherenkov
detector\/}\@.
The photons were measured by an {\em elctromagnetic calorimeter\/} made
of CSI(Tl) crystals.
The magnetic field of 1.5 T was generated by a superconducting solenoid.
The iron of the flux return was instrumented by RPCs and LSTs for measuring the
muons.

\section{Data collection}
\mbox{~~~~The data} used in this analysis include 288 ${\rm fb}^{-1}$ taken at
the $\Upsilon(4S)$ resonance and 27 ${\rm fb}^{-1}$ collected below the 
resonance.

The cuts applied for selecting the $D^0 \rightarrow \pi^+ \pi^- \pi^0$ 
candidates are \cite{Mishra}:
(i)    charged tracks with $p_T > 100$ MeV/$c$;
(ii)   particle identification compatibile with a charged pion;
(iii)  $E_\gamma > 100$ MeV;
(iv)   $115 < M(\gamma \gamma) < 150$ MeV/$c^2$;
(v)    $E(\gamma \gamma) > 350$ MeV;
(vi)   $D^0$ vertex fit with $P(\chi^2) > 0.5\%$;
(vii)  $1848 < M(D^0) < 1880$ MeV/$c^2$;
(viii) $D^0$ generated by the $D^{*+}$ decay and selected with the cut
       $|M(D^{*+}) - M(D^0) - 145.4| < 0.6$ MeV/$c^2$;
(ix)   $D^0$ momentum in the $e^+ e^-$ c.m.s.\ $p^* > 2.77$ MeV/$c$;
(x)    exclusion of the events with $489 < M(\pi^+ \pi^-) < 508$ MeV/$c^2$.

\begin{figure}
\includegraphics[height=0.5\textheight]{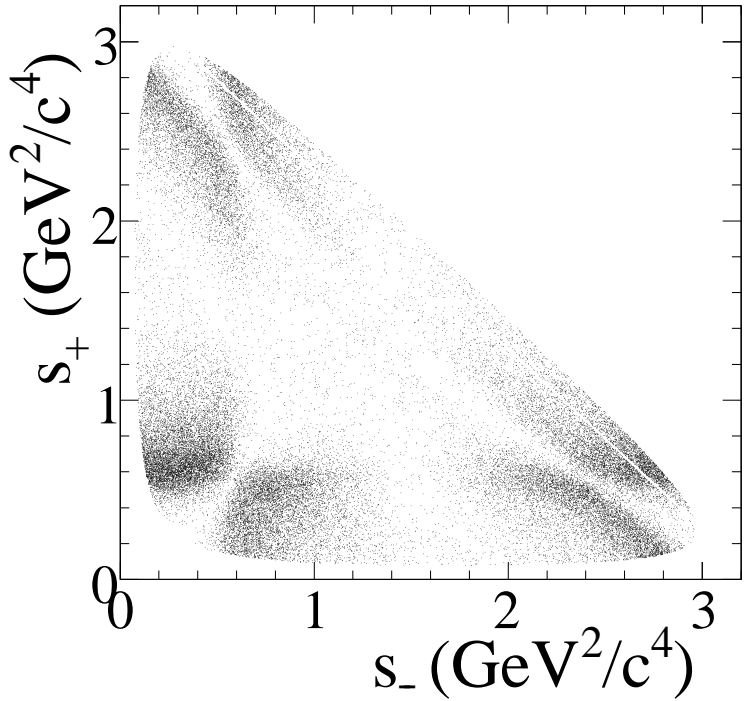}
\caption{
The Dalitz-plot of the $D^0 \rightarrow \pi^+ \pi^- \pi^0$ decay.
$s_+$ and $s_-$ are respectively $m^2(\pi^+ \pi^0)$ and $m^2(\pi^- \pi^0)$\@.
The fine diagonal line at low $\pi^+ \pi^-$ mass corresponds to the events 
removed by the cut $489 < M(\pi^+ \pi^-) < 508$ MeV/$c^2$.
The Dalitz-plot shows three diagonals with low density, indicating the 
dominance of the isospin zero.
}
\end{figure}

The Dalitz-plot of the events selected with these cuts is shown in Fig.~1\@.
It was already published in Refs. \cite{BABAR07,GMMS}\@.
It contains $44\ 780 \pm 250$ $D^0$ decays and an estimated contamination of 
$830 \pm 250$ events.
The contamination was evaluated using the sideband 
$1930 < M(\pi^+ \pi^-\pi^0) < 1990$ MeV/$c^2$.

This Dalitz-plot shows three $\rho$ bands and has low density at the centre and
on the three diagonals.
This structure is typical of a $\pi^+ \pi^- \pi^0$ final state with 
$I^G J^P = 0^- 0^-$ \cite{Zemach}.
The same properties were also visible in the CLEO analysis of the same decay
\cite{CLEO05}.

\section{Dalitz-plot analysis}
\mbox{~~~~The Dalitz} plot density was fitted with the ansatz
\begin{equation}
D(s_+,s_-) = {\cal N} |a_{\rm NR} e^{i \phi_{\rm NR}} + \sum_n a_n e^{i \phi_n}
A_n(s_+ s_-)|^2 ,
\label{density}
\end{equation}
where $s_+ = m^2(\pi^+ \pi^0)$, $s_- = m^2(\pi^- \pi^0)$,
$\cal N$ is the normalization factor such that 
$\int D(s_+,s_-) \, ds_+ ds_- = 1$, and $A_n(s_+,s_-)$ is the amplitude for 
the $n$.th channel.

The amplitude for the decay $D^0 \rightarrow R_n \pi_3$, 
$R_n \rightarrow \pi_1 \pi_2$ was calculated as
\begin{equation}
A_n(s_+ s_-) = \frac{h_n S_J}{m^2_n - m^2_{12} - i m_n \Gamma_n(m_{12})} ,
\label{amplitude}
\end{equation}
where $h_n$ is a normalization factor evaluated such that 
$\int |A_n(s_+ s_-)|^2 \, ds_+ ds_- = 1$,
$m_n$ is the mass of the resonance $R_n$,
$S_J$ is the spin factor for a resonance with spin $J$,
and $\Gamma_n(m_{12})$ is the variable width of the resonance.

The functions $S_J$ and $\Gamma_n(m_{12})$ were written using the formulae 
written by the CLEO Collaboration in the analysis of the decay 
$D^0 \rightarrow K^- \pi^+ \pi^0$ \cite{CLEO01}

The result of the fit is reported in Table~I\@.
It shows that the $D^0$ Dalitz-plot is dominated by the three $\rho(770) \pi$
channels, with a small contribution of the three $\rho(1700) \pi$ channels and
a very small contribution of the other channels.
Furthermore, the sum of the fractions is 147.4\%\@.
This fact indicates that there is a strong negative interference between the
16 amplitudes of the channels used in the analysis.

\begin{table}
\begin{tabular}{lccc}
\hline
Channel                &    Amplitude $a_n$       & Phase $\phi_n$ $(^\circ)$ &
 Fraction $f_n$ (\%)  \\
\hline
$\rho(770)^+ \pi^-$    & $0.823 \pm 0.000 \pm 0.004$ &        0          &
       $67.8 \pm 0.0 \pm 0.6$ \\
$\rho(770)^0 \pi^0$    & $0.512 \pm 0.005 \pm 0.011$ & $16.2 \pm 0.6 \pm 0.4$ &
       $26.2 \pm 0.5 \pm 1.1$ \\
$\rho(770)^- \pi^+$    & $0.588 \pm 0.007 \pm 0.003$ & $-2.0 \pm 0.6 \pm 0.6$ &
       $34.6 \pm 0.8 \pm 0.3 $ \\
$\rho(1450)^+ \pi^-$   & $0.033 \pm 0.011 \pm 0.018$ & $-146 \pm 18 \pm 14~$ &
       $0.11 \pm 0.07 \pm 0.12$ \\
$\rho(1450)^0 \pi^0$   & $0.055 \pm 0.010 \pm 0.006$ & $~~~10 \pm 8 \pm 13$   &
       $0.30 \pm 0.11 \pm 0.07$ \\
$\rho(1450)^- \pi^+$   & $0.134 \pm 0.008 \pm 0.004$ & $~~16 \pm 3\pm 3$     &
       $1.79 \pm 0.22 \pm 0.12$ \\
$\rho(1700)^+ \pi^-$   & $0.202 \pm 0.017 \pm 0.017$ & $-17 \pm 2 \pm 2$    &
       $4.1 \pm 0.7 \pm 0.7$ \\
$\rho(1700)^0 \pi^0$   & $0.224 \pm 0.013 \pm 0.022$ & $-17 \pm 2 \pm 3$    &
       $5.0 \pm 0.6 \pm 1.0$ \\
$\rho(1700)^- \pi^+$   & $0.179 \pm 0.011 \pm 0.017$ & $-50 \pm 3 \pm 3$    &
       $3.2 \pm 0.4 \pm 0.6$ \\
\hline
$f_0(400) \pi^0$       & $0.091 \pm 0.006 \pm 0.006$ & $~~~~8 \pm 4 \pm 8$    &
       $0.82 \pm 0.10 \pm 0.10$ \\
$f_0(980) \pi^0$       & $0.050 \pm 0.004 \pm 0.004$ & $-59 \pm 5 \pm 4$    &
       $0.25 \pm 0.04 \pm 0.04$ \\
$f_0(1370) \pi^0$      & $0.061 \pm 0.009 \pm 0.007$ & $156 \pm 9 \pm 6$    &
       $0.37 \pm 0.11 \pm 0.09$ \\
$f_0(1500) \pi^0$      & $0.062 \pm 0.006 \pm 0.006$ & $~~12 \pm 9 \pm 4$    &
       $0.39 \pm 0.08 \pm 0.07 $ \\
$f_0(1710) \pi^0$      & $0.056 \pm 0.006 \pm 0.007$ & $~~51 \pm 8 \pm 7$    &
       $0.71 \pm 0.07 \pm 0.08$ \\
$f_2(1270) \pi^0$      & $0.115 \pm 0.003 \pm 0.004$ & $-171 \pm 3 \pm 4~~$  &
       $1.32 \pm 0.08 \pm 0.10$ \\
\hline
nonresonant            & $0.092 \pm 0.011 \pm 0.007$ & $-11 \pm 4 \pm 2$ &
       $0.84 \pm 0.21 \pm 0.12$ \\
\hline
\end{tabular}
\caption{
Result of the fit of the $D^0$ Dalitz-plot showing the amplitude $a_n$, the 
phase $\phi_n$ and the fraction $f_n = a^2_n$\@.
The mass and width of the $f_0(400)$ are 400 and 600 MeV/$c^2$\@.
The masses and widths of the other mesons are taken by the 2006 issue of the
{\em Review of Particle Physics\/} [8].
The phase of the $\rho(770)^+ \pi^-$ is fixed at $0^\circ$.
}
\end{table}

\section{Isospin decomposition}

\mbox{~~~~The first nine} channels reported in Table~I have 
the pions 1 and 2 in the isospin eigenstate $I_{12} = 1$,
the subsequent six channel  $I_{12} = 0$, and the last, i.e.\ the 
{\em non resonant} channel, is not an eigenstate of $I_{12}$\@.
The Feynman diagrams of these channels are shown in Fig.~2\@.
They can be grouped into four channels

\begin{eqnarray}
|\{\rho\}^+ \pi^- \rangle & = & |\rho(770)^+ \pi^- \rangle + \, 
|\rho(1450)^+ \pi^- \rangle + 
\, |\rho(1700)^+ \pi^- \rangle \, = \nonumber \\
&  &\frac{\displaystyle 1}{\displaystyle \sqrt{2}} \, 
( |+0- \rangle - \, |0+- \rangle ) , \label{Idipion}  \\
|\{\rho\}^0 \, \pi^0 \rangle~ & = & |\rho(770)^0 \pi^0 \rangle + \, 
|\rho(1450)^0 \pi^0 \rangle + 
\, |\rho(1700)^0 \pi^0 \rangle \, = \nonumber \\
& &\frac{\displaystyle 1}{\displaystyle \sqrt{2}} \, 
( |-+0 \rangle - \, |+-0 \rangle ) , \nonumber \\
|\{\rho\}^- \pi^+ \rangle & = & |\rho(770)^- \pi^+ \rangle + \, 
|\rho(1450)^- \pi^+ \rangle + \,
|\rho(1700)^- \pi^+ \rangle \, = \nonumber \\
& &\frac{\displaystyle 1}{\displaystyle \sqrt{2}} \, 
( |-0+ \rangle - \, |0-+ \rangle ) , \nonumber \\
|\{f\} \ \pi^0 \rangle~ & = & |f_0(400) \pi^0 \rangle + \, 
|f_0(980) \pi^0 \rangle + 
\, |f_0(1370) \pi^0 \rangle + \nonumber \\[5pt]
& & |f_0(1500) \pi^0 \rangle + \, |f_0(1710) \pi^0 \rangle + \, 
|f_2(1710) \pi^0 \rangle \, =
\nonumber \\
& &  \frac{\displaystyle 1}{\displaystyle \sqrt{2}} \, ( |+-0 \rangle + \, 
|-+0 \rangle ) , 
\nonumber
\end{eqnarray}
where $|q_1 q_2 q_3 \rangle$ indicates the state 
$|\pi^{q_1} \rangle \, |\pi^{q_2} \rangle \, |\pi^{q_3} \rangle$,
$q_i = +, -, 0$ being the charge of the $i.$th pion.
The amplitudes of these four states are obtained by summing the amplitudes of
the channels in the right side of \eqref{Idipion}.

\begin{figure}
\includegraphics[height=0.35\textheight]{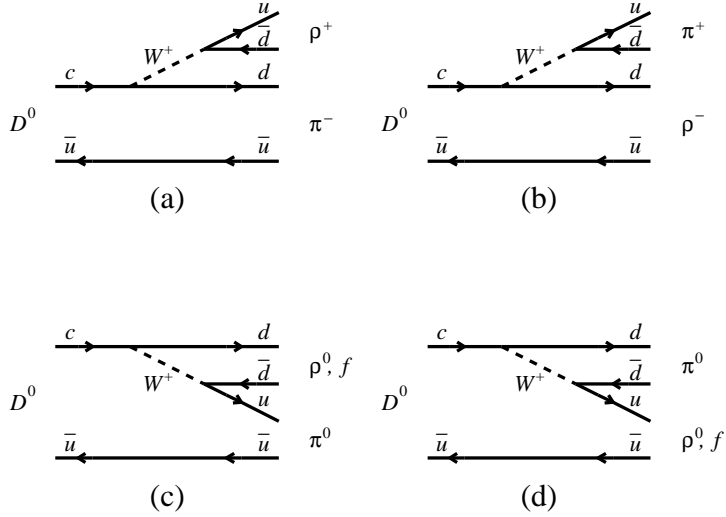}
\caption{
The Feynman diagrams for the first fitheen channels listed in Table~I\@. 
(a) $|\{\rho\}^+ \pi^- \rangle$\@.
(b) $|\{\rho^-\} \pi^+ \rangle$\@.
(c) and (d) $|\{\rho^0\} \pi^0 \rangle$ and $|\{f\} \pi^0 \rangle$.
$\{\rho\}$ indicates one of the three mesons $\rho(770)$, $\rho(1450)$, and
$\rho(1700)$\@.
$\{f\}$ indicates one of the six mesons $f_0(400)$, $f_0(980)$, $f_0(1370)$,
$f_0(1500)$, $f_0(1710)$, and $f_2(1270)$.
}
\end{figure}

We use the symbols $|I(I_{12});I_z \rangle$ for the isospin eigenstates of the
$3 \pi$ final states.
Here $I$ is the total isospin, $I_{12}$ is the isospin of interacting pion 
pair 12, and $I_z$ is the third component of the isospin.
The formulae of these eigenstates with $I_z = 0$ can be found in Eq.\ (3) of
Ref.\ \cite{GMMS}.

Using the relations \eqref{Idipion}, the three eigenstates with 
$I_{12} = 1$ can be written such a way
\begin{eqnarray}
|2(1);0 \rangle & = & \frac{\displaystyle 1}{\displaystyle \sqrt{6}} \,
\left( |\{\rho\}^+ \pi^- \rangle - 2 |\{\rho\}^0 \pi^0 \rangle + 
|\{\rho\}^- \pi^+ \rangle \right) 
, \label{stati1} \\
|1(1);0 \rangle & = & \frac{\displaystyle 1}{\displaystyle \sqrt{2}} \,
\left( |\{\rho\}^+ \pi^- \rangle - |\{\rho\}^- \pi^+ \rangle \right) , 
\nonumber \\
|0(1);0 \rangle & = & \frac{\displaystyle 1}{\displaystyle \sqrt{3}} \,
\left( |\{\rho\}^+ \pi^- \rangle +  |\{\rho\}^0 \pi^0 \rangle + 
|\{\rho\}^- \pi^+ \rangle \right) . \nonumber 
\end{eqnarray}

The {\em non resonant\/} channel $|{\rm NR} \rangle$ can have two 
interpretations:
(i)  it corrects the $|\{f\} \pi^0 \rangle$ amplitude that was not well 
parametrized;
(ii) it describes a point-like interaction that generates a uniform 
$\pi^+\pi^- \pi^0$ final state.
In the case (i), the channel $|{\rm NR} \rangle$ has $I_{12} = 0$\@.
Therefore, the contribution of the isospin state $|1(0);0 \rangle$ to the 
$D^0 \rightarrow \pi^+ \pi^- \pi^0$ decay is
\[ P_{\{+-0\}}|1(0);0 \rangle = |\{f\} \pi^0 \rangle + |{\rm NR} \rangle , \]
$P_{\{+-0\}}$ being the projection operator of a isospin eigenfunction into
the final state $\pi^+ \pi^- \pi^0$.

In the case (ii), the isospin wave functions of $|{\rm NR} \rangle$ is 
symmetrical.
There are two $3 \pi$ symmetrical isospin eigenstates.
One is$|3(2);0 \rangle$, the other is the $I=1$ symmetric state
\begin{eqnarray*} 
|1(S);0 \rangle & \equiv & \frac{2}{3} \, |1(2);0 \rangle + 
\frac{\sqrt{5}}{3} \, |1(0);0 \rangle 
 =  \frac{1}{\sqrt{15}} \, ( |+0- \rangle + |+-0 \rangle + \\
& &  |0+- \rangle + |0-+ \rangle + |-+0 \rangle + |-0+ \rangle - 
3 |000 \rangle ) .
\end{eqnarray*}

The analysis carried out in Ref.\ \cite{GMMS} was based on the assumption that
the isospin wave-function of the {\em non resonant\/} channels was 
$|1(S);0 \rangle$, because a state generated by a point-like four quark 
interaction cannot have $I=3$\@.
This interpretation predicts
\begin{eqnarray*}
 P_{\{+-0\}}|1(0);0 \rangle & = & |\{f\} \pi^0 \rangle +  
\frac{\sqrt{5}}{3} \, |{\rm NR} \rangle  , \\
 P_{\{+-0\}}|1(2);0 \rangle & = & \frac{2}{3} \, |NR \rangle .
\end{eqnarray*}

\begin{table}
\begin{tabular}{lccc}
\hline
Isospin wave function  &    Amplitude  & Phase $(^\circ)$ &  Fraction (\%)  \\
\hline
$|2(1);0 \rangle$  
              & $0.1368 \pm 0.0016$ & $-42.5 \pm 0.7~~$ &  $1.87 \pm 0.04$ \\
$|1(2);0 \rangle$  
              & $0.0617 \pm 0.0022$ &  $-8.9 \pm 2.6$   &  $0.38 \pm 0.03$ \\
$|1(1);0 \rangle$  
              & $0.0799 \pm 0.0023$ &  $18.0 \pm 2.0$   &  $0.64 \pm 0.04$ \\
$|1(0);0 \rangle$  
              & $0.0936 \pm 0.0051$ &  $14.5 \pm 2.4$   &  $0.87 \pm 0.10$ \\
$|0(1);0 \rangle$  
              & $0.9810 \pm 0.0006$ &         0         & $96.23 \pm 0.12~~$ \\
\hline
\end{tabular}
\caption{
Amplitudes, phases and fractions of the five isospin channels contributing to 
the $D^0 \rightarrow \pi^+ \pi^- \pi^0$ decay.
The errors are only statistical.
The phase of the state $|0(1);0 \rangle$ is fixed at $0^\circ$.
}
\end{table}

The results are reported in Table~II\@.

These results allow to estimate the branching ratio 
${\cal B} (D^0 \rightarrow 3 \pi^0)$\@.
The branching ratio of the decay \eqref{D03pi} is
${\cal B} (D^0 \rightarrow \pi^+ \pi^- \pi^0) = (1.44 \pm 0.06)\%$ \cite{PDG08}
and Eq.\ (3) of Ref.\ \cite{GMMS} tell us that the isospin eigenstates
$|1(2);0 \rangle$ and $|1(0);0 \rangle$ decays into $3 \pi^0$ and 
$\pi^+ \pi^-\pi^0$ 
respectively with the ratios 4:11 and 1:2\@.
Therefore, from the fractions $f$ reported in the last coluumn of Table~II, we
obtain
\begin{eqnarray*} {\cal B} (D^0 \rightarrow 3 \pi^0) & = &
{\cal B} (D^0 \rightarrow \pi^+ \pi^- \pi^0) 
\left[ \frac{4}{11} f_{|1(2);0 \rangle} +
\frac{1}{2} f_{|1(0);0 \rangle} \right] = \\
& & (8.3 \pm 0.8) \times 10^{-5} .
\end{eqnarray*}

This estimate is in agreement with the measure of CLEO 
${\cal B} (D^0 \rightarrow 3 \pi^0) < 3.5 \times 10^{-4}$ \cite{CLEO06}\@.

\section{Interpretation and predictions}
\mbox{~~~~A $\pi^+ \pi^- \pi^0$} final state generated by a pseudoscalar meson
decay has $J^P = 0^-$, $G$-parity $-1$, and charge conjugation $C = G (-1)^I$.
The isospin states $|0(1);0 \rangle$ and $2(1);0 \rangle$ have $CP = +1$ and 
the other three states with $I = 1$ have $CP = -1$\@.
Therefore, the results shown in Table~II tell us that the
decay $D^0 \rightarrow \pi^+ \pi^- \pi^0$ proceeds for $(98.11 \pm 0.11)\%$
via the $CP = +1$ eigenstate
\[ D_1 = \frac{1}{\sqrt{2}} ( |D^0 \rangle + |\bar{D}^0 \rangle ) . \]

The tree graphs shown in Fig.~2 {\em do not\/} predict the dominance of a pure
isospin state.
Then, if the $I = 0$ dominance is not coincidental, {\em there should be a 
physical explanation.\/}
A possible interpretation is that the $I = 0$ dominance is due to a final 
state interaction with an $I^G J^{PC} = 0^- 0^{--}$ meson that 
resonates with the four quark generated by the $D^0$ decay.
Such a meson must be exotic because a $q \bar{q}$ pair cannot have these 
quantum number.
It could be a state $2 q 2 \bar{q}$\@.

If this interpretation is right, this meson should also have other 
decays.
The pricipal candidates are the final states $2 \pi^+ 2 \pi^- \pi^0$ and
$\pi^+ \pi^- 3 \pi^0$ because the analysis of the 
$D^0 \rightarrow \pi^+ \pi^- \pi^0$ found a non negligible contributions 
of the channels $\rho(1450) \pi$ and $\rho(1700) \pi$, and the mesons 
$\rho(1450)$ and $\rho(1700)$ decay also in $4 \pi$\@.
Furthermore, it is possible that this meson could also decay into 
$K \bar{K} \pi$.

\section{Conclusions}
\mbox{~~~~We see a strange} behaviour in the 
$D^0 \rightarrow \pi^+ \pi^- \pi^0$ Dalitz-plot: 
this decay is dominated by the $I = 0$ final state \cite{BABAR07,GMMS}\@.
We take this as a hint of someting interesting that deserves further studies.
We need to analyze the decays $D^0 \rightarrow K \bar{K} \pi$ and 
$D^0 \rightarrow 5 \pi$ to understand if the $I = 0$ dominance in 
$D^0 \rightarrow \pi^+ \pi^- \pi^0$ is coincidental or it is a general rule not
predicted by the standard model.

\section*{Acknowledgments}
\mbox{~~~~I would} like my colleagues Brian Meadows, Kalahand Mishra and Abi 
Soffer that have carried out togheter with me the analysis of the $I = 0$ 
dominance and Dr.\ Fabio Ferrarotto who helped me to write this manuscript.

\end{document}